\documentclass[a4paper]{aa}

\usepackage{graphicx}
\usepackage{natbib}
\bibpunct{(}{)}{;}{a}{}{,}

\usepackage{txfonts}

\begin{document}
\title{Formation of giant planets around stars with various masses}
%\title{}
\author{K. Kornet\inst{1,2} \and S. Wolf\inst{1}
\and M. R{\'o}{\.z}yczka\inst{2}}
\institute{Max Planck Institute for
Astronomy, K\"onigstuhl 17, 69117 Heidelberg, Germany
\and %
Nicolaus Copernicus Astronomical Center, Bartycka 18,
Warsaw, 00-716, Poland }

\abstract{We examine the predictions of the core accretion - gas
capture model concerning the efficiency of planet formation around
stars with various masses. First, we follow the evolution of gas
and solids from the moment when all solids are in the form of
small grains to the stage when most of them are in the form of
planetesimals. We show that the surface density of the
planetesimal swarm tends to be higher around less massive stars.
Then, we derive the minimum surface density of the planetesimal
swarm required for the formation of a giant planet both in a
numerical and in an approximate analytical approach. We combine
these results by calculating a set of representative disk models
characterized by different masses, sizes, and metallicities, and
by estimating their capability of forming giant planets. Our results
show that the set of protoplanetary disks capable of giant planet
formation is larger for less massive stars. Provided that the
distribution of initial disk parameters does not depend too
strongly on the mass of the central star, we predict that the
percentage of stars with giant planets should increase with
decreasing stellar mass. Furthermore, we identify the 
radial redistribution of solids during the formation of
planetesimal swarms as the key element in explaining these effects.}

\maketitle

\section{Introduction}

Radial velocity surveys led to the discovery of over 150
extrasolar planets around main sequence stars. Published
descriptions of most of them can be found in the references
given by \citet{marcy05} and \citet{mayor04}. Those surveys have
been the most successful in the case of G dwarf stars, because
such stars have well-defined spectroscopic features and show only
a little photospheric activity. Consequently, most of the known
extrasolar planets orbit stars similar to our Sun. Due to the
constant progress in the detection techniques, the observational
programs recently started to also include stars with lower masses
on a larger scale, namely M dwarfs. Moreover, some of these
surveys are now particularly dedicated to lower-mass stars
\citep[e.q.][]{Endl03,Bonfils04}. So far, these efforts have led to the
discovery of three planets around two M dwarf stars: Gliese 876b,c
\citep{marcy98,marcy01} and GJ436b \citep{butler04}.

From the theoretical point of view, the problem of giant planet
formation around M dwarfs was studied recently by
\citet{laughlin04}. They addressed it within the core accretion -
gas capture model (CAGCM) that provides the most widely accepted
scenario explaining the formation of giant planets in both the
Solar System and extrasolar planetary systems. This model predicts
that first a solid planetary core is formed by collisional
accumulation of planetesimals. When the core reaches a mass of a few
Earth masses, it starts to accrete
gas, and an extended hydrostatic envelope is built around it.  As
the accretion rate of gas is greater than the
accretion rate of solids at this time, the envelope eventually becomes more
massive than the core. When this happens, a runaway accretion of
gas ensues, which is terminated either by tidal interactions of
the planet with the protoplanetary disk or by the dissipation of
the disk. CAGCM has found supporting evidence in the discovery
that stars with planets have higher metallicities than field stars
\citep{santos00,debra03}. This is because the formation time of
giant planets decreases with increasing surface density of the
planetesimal swarm \citep{pollack96}, which in turn increases with
the primordial metallicity of the protoplanetary disk. Thus, giant
planets are expected to form more easily in disks with higher
metallicity.

\citet{laughlin04} conclude that M dwarfs have a limited ability
to form Jupiter-mass planets. This is a direct consequence of
their assumption that the surface density of the planetesimal
swarm out of which planetary cores are formed scales linearly with
the mass of the central star. However, the solid component of the
protoplanetary disk evolves in a different way than the gaseous
component \citep{weiden93}. Due to the gas drag, a significant
redistribution of solids takes place, and in the inner disk their
surface density can be substantially enhanced compared to the
initial one \citep{weiden03,SV97}. In general, the efficiency of
the processes responsible for the redistribution of dust depends
on the mass of the central star. An obvious conclusion is that 
analysis of the formation of giant planets around stars with
various masses should include the global evolution of solids in
protoplanetary disks. A simple model of the evolution of solids
was proposed by \citet{kac2,kac05}. Applying it to
solar-like central stars, these authors reproduced
the observed correlation between stellar metallicity and
the probability of a planet occurring \citep[a similar result was
independently obtained by][]{ida04b}.

The rapid progress in observational techniques opens up the possibility
of testing the correctness and predictive power of the model proposed
by \citet{kac05}. To that end, we extend their analysis and
calculate probabilities of planet occurrence around stars with
different masses (both smaller and larger than $1 M_\odot$), which
may be compared to future observational data. In Sect.
\ref{s:methods} we explain our approach to the evolution of
protoplanetary disks and planet formation. The results of our
calculations are presented in Sect. \ref{s:res} and discussed in
Sect. \ref{s:conc}.

\section{Methods of calculation}
\label{s:methods}
\subsection{The disk}

We model the protoplanetary disk as a two-component fluid,
consisting of gas and solids. The gaseous component is described
by the analytical model of \citet{s98}, which gives the surface
density of gas, $\Sigma_\mathrm{g}$, as a function of distance $a$
from the star  and time $t$, in terms of a selfsimilar solution to
the viscous diffusion equation. The viscosity coefficient is given
by the standard $\alpha$ prescription:
\begin{equation}
\nu=\frac{1}{3} \alpha C_\mathrm{S} H
\end{equation}
where $H$ is a density scale-height of gaseous disk and $C_\mathrm{S}$ denotes the
 speed of sound in the gas. All other quantities characterizing the gas are
 obtained in a thin disk and vertical thermal balance approximation by
 solving the set of equations \citep[see, for example][]{accr_power}:
\begin{eqnarray}
  \label{eq:th-d1}
   \Sigma_\mathrm{g}& = &2 H \rho_\mathrm{g} \\
   H& = &\frac{\sqrt{2} C_\mathrm{s}}{\Omega_\mathrm{K}}\\
   C_\mathrm{s}^2  & = &\frac{k_\mathrm{B} T}{\mu m_\mathrm{H}}\\
   \label{eq:th-dprzedost}
   \frac{16 \sigma_\mathrm{B} T^4}{3 \kappa \Sigma_\mathrm{g}}& = &\frac{9}{4} \Sigma_\mathrm{g} \nu
   \Omega_\mathrm{K}^2 \quad .
   \label{eq:th-dost}
\end{eqnarray}
Here $T$ is the temperature in the midplane of the disk, and
$\Omega_\mathrm{k}$ the Keplerian angular velocity. For the Rosseland
mean opacity $\kappa$ the analytical piecewise-continuous power law
formulas from \citet{ruden91} were adopted.  In the parts of the disk
where the optical thickness falls below a critical value of
$\tau_\mathrm{crit}=1.78$ \citep{ruden91}, the last equation is
replaced by:
\begin{equation}
\frac{4 \sigma_\mathrm{B} \kappa T^4}{3 \tau_\mathrm{crit}} =
\frac{9}{4} \nu    \Omega_\mathrm{K}^2
\end{equation}

The main assumptions underlying our approach to the evolution of
solids are the folllowing. (1) At each radial distance from the central star the
particles have the same size (which in general varies over~ time).
(2) There is only one component of dust, in this case
corresponding to high-temperature silicates with the evaporation
temperature $T_\mathrm{evap}$ = 1350 K and bulk density 3.3 g
cm$^{-3}$. This choice is justified by the fact that in most
cases the surface density of a water-ice planetesimal swarm would
be too low in the range of distances considered here, e.g. $r\le5$
AU, to enable formation of a giant planet in a reasonable time
\citep[see also][]{kac4}. We include other species, e.g. water ice
only, in modeling the gaseous disk as the sources of opacity. In
other words we assume that the evolution of silicate grains is
not strongly influenced by the evolution of i.e. ice grains. (3)
All collisions between particles lead to coagulation. (4) When the
temperature exceeds $T_\mathrm{evap}$, local solids immediately
sublimate and the vapour evolves at the same radial velocity as
the gas component. (5) Initially, in regions where the disk
temperature is lower than $T_\mathrm{evap}$, all solids are in the
form of grains with radii $10^{-3}\ \mathrm{cm}$ (the results do
not depend on the choice of that particular value, as long as the
solids are initially small enough to be coupled well to the gas).
(6) The radial velocities of solid particles are entirely
determined by the effects of gas drag. (7) The relative velocities
of solid particles when they collide are computed according to
the turbulent model described by \citet{SV97}.  (8) The
evolution of the solids does not affect the evolution of the gas.
At each radius, the vertical extent of the solid particle
distribution is calculated and is evolved in time, so the effect
of the sedimentation of solids toward the midplane of the disk is
taken into account.  All assumptions and approximations are
discussed in \citet{SV97}, \citet{s98}, and \citet{kac1}.

The evolution of solids is governed by two equations. The first of
them is the continuity equation for the surface density of solid
material, $\Sigma_\mathrm{s}$. The second one, describing
the evolution of grain sizes, can be interpreted as the continuity
equation for size-weighted surface density of solids,
$\Sigma_\mathrm{size}\equiv s(a)\Sigma_\mathrm{s}$, where $s(a)$
is the radius of solid particles at a distance $a$ from the star.
The equations are solved numerically on a moving grid whose outer
edge follows the outer edge of the dust disk. The details of the
method can be found in \citet{kac1}.

\subsection{The planets}
We model the formation of a giant planet \textit{in situ}, so the
orbital parameters of the planet do not vary in time. Our
procedure for the evolution of the protoplanetary cores is based
on the following assumptions: (1) core accretion starts when
solids at a given distance $a$ from the star reach  radii of $2\
\mbox{km}$; (2) at each time, the planetesimals are mixed
well through the feeding zone of the planet, so their surface
density $\Sigma_\mathrm{s}$ is  always uniform in space, but
usually decreasing with time as pla\-ne\-te\-simals accrete onto
the planet; (3) the planetesimals do not migrate into the feeding
zone from outside or inside and vice versa, but they can be
overtaken by the boundary of the feeding zone as it expands due to
the growing mass of the planet. Under these assumptions the growth
of protoplanetary core mass $M_\mathrm{c}$ can be described by
the formula given by \citet{papaloizou99},
\begin{equation}
\dot{M_\mathrm{c}}=C_1 C_\mathrm{cap} R_p R_H \Omega_\mathrm{K} 
\Sigma_\mathrm{s} \quad ,
\label{eq:mc}
\end{equation}
where
\begin{equation}
R_\mathrm{H}=a\left(\frac{M_\mathrm{p}}{h M_\star}\right)^{1/3}
\end{equation}
is the radius of the Hill sphere of the planet. The quantity $h$
is a constant factor that reflects different definitions of the
Hill radius in the literature. Herein we assume $h=3$. The value
of $C_1$ given by \citet{papaloizou99} is $81\pi/32$; we use a
factor of 5 ( the difference comes from a different definition
of $R_\mathrm{H}$). The quantity $C_\mathrm{cap}$ describes the
increase in the effective capture radius of the planet with respect
to its real radius $R_\mathrm{p}$ due to interaction of
planetesimals with the envelope of the planet \citep{podolak88}.
We approximate it with a fit to the results of \citet{boden00}
provided by \citet{hubickyj01}. For core masses less than $5
M_{\odot}$, no increase in the effective capture radius is assumed,
i.e. $C_\mathrm{cap}(M_\mathrm{c}<5M_\oplus)=1$. For higher core
masses, it increases linearly  with the mass of the core, reaching
its maximum value of $C_\mathrm{cap}=5$ for $M_\mathrm{c}=15
M_\oplus$. The surface density of planetesimals
$\Sigma_\mathrm{s}$ also changes in time, as they are accreted
onto the core, and the feeding zone expands. Under our assumptions
it can be calculated as
\begin{equation}
\Sigma_\mathrm{s}=\Sigma_\mathrm{s,init}-\frac{M_\mathrm{c}}{2\pi a \Delta a}
\label{eq:sigmad}
\end{equation}
where $\Sigma_\mathrm{s,init}$ is the initial surface density of planetesimals,
and $\Delta a$ is the width of the feeding zone. Herein we assume that
$\Delta a=8 R_\mathrm{H}$ \citep{sf_odz}.

To calculate the rate of gas accretion onto the protoplanet,  one should  
solve the equations of mass conservation,
hydrostatic equilibrium, energy generation from accretion of
planetesimals and quasi-static contraction, and radiative or
convective energy transport, as given e.g. in \citet{boden86}.
 Following \cite{ida04a}, we use a simplified approach based
on fits to the numerical results. We assume that the accretion of
gas starts when the core reaches critical mass of
\begin{equation}
M_\mathrm{c,crit}= 10 \left(\frac{\dot{M}_\mathrm{c}}{10^{-6}
  \ M_\oplus\ yr^{-1}}\right)^{0.2-0.3}
\left(\frac{\kappa_\mathrm{env}}{1\ \mathrm{cm}^2\ \mathrm{g}^{-1}}\right)^{0.2-0.3}
\label{e:mcrit}
\end{equation}
\citep{ikoma00}, where $\kappa_\mathrm{env}$ is the opacity in the
envelope of the planet. Its actual magnitude is currently poorly
constrained. We assume that $\kappa_\mathrm{env}=1\ \mathrm{cm}^2\
\mathrm{g}^{-1}$. For the value of the powerlaw index in  the
dependence on $M_\mathrm{c}$ (Eq. \ref{e:mcrit}) we adopt $0.25$.
When the mass of the protoplanet is higher than
$M_\mathrm{c,crit}$ it  contracts on the Kelvin-Helmholtz time
scale $\tau_\mathrm{KH}$. \citet{bryden00} show by fitting the
result of \citet{pollack96} that
\begin{equation}
\tau_\mathrm{KH}=10^b \left(\frac{M_\mathrm{p}}{M_\oplus}\right)^{-c+1}
    \left(\frac{\kappa}{1 \mathrm{cm^2\ g^{-1}}}\right)\ \mathrm{yr}
\end{equation}
where $b=10$ and $c=-4$. Consequently, for
$M_\mathrm{p}>M_\mathrm{c,crit}$ we adopt the following
equation for the gas accretion rate onto the planet
\begin{equation}
\frac{\mathrm{d} M_\mathrm{env}}{\mathrm{d}t} =
  \frac{M_\mathrm{p}}{\tau_\mathrm{KH}} = A
  \left(\frac{M_\mathrm{p}}{M_\oplus}\right)^{-c}\ M_\oplus
  \ \mathrm{yr}^{-1} \equiv \tilde{A} M_\mathrm{p}^{-c},
\label{eq:menv}
\end{equation}
where $A=10^{-10}$. The free parameters in Eqs. (\ref{eq:mc} -- \ref{eq:menv}) are the
mass of the central star $M_\star$, the distance $a$ of
protoplanet from the star, and the initial surface density of
planetesimals $\Sigma_\mathrm{s,init}$.

\section{Results}
\label{s:res}
In the CAGCM the process of planet formation is naturally split into two
main phases. In the first one, dominated by collisional
accumulation of dust grains, a planetesimal swarm is formed in the
protoplanetary disk. In the second phase, dominated by
gravitational interactions, planetary cores are assembled and 
subsequently accrete planetesimals and gas from the disk. In the
following subsections we investigate how each phase is influenced
by the mass of the central star.

\subsection{From dust grains to planetesimals}
\label{s:exam}
\begin{figure*}
\centering
\includegraphics[angle=-90,width=8cm]{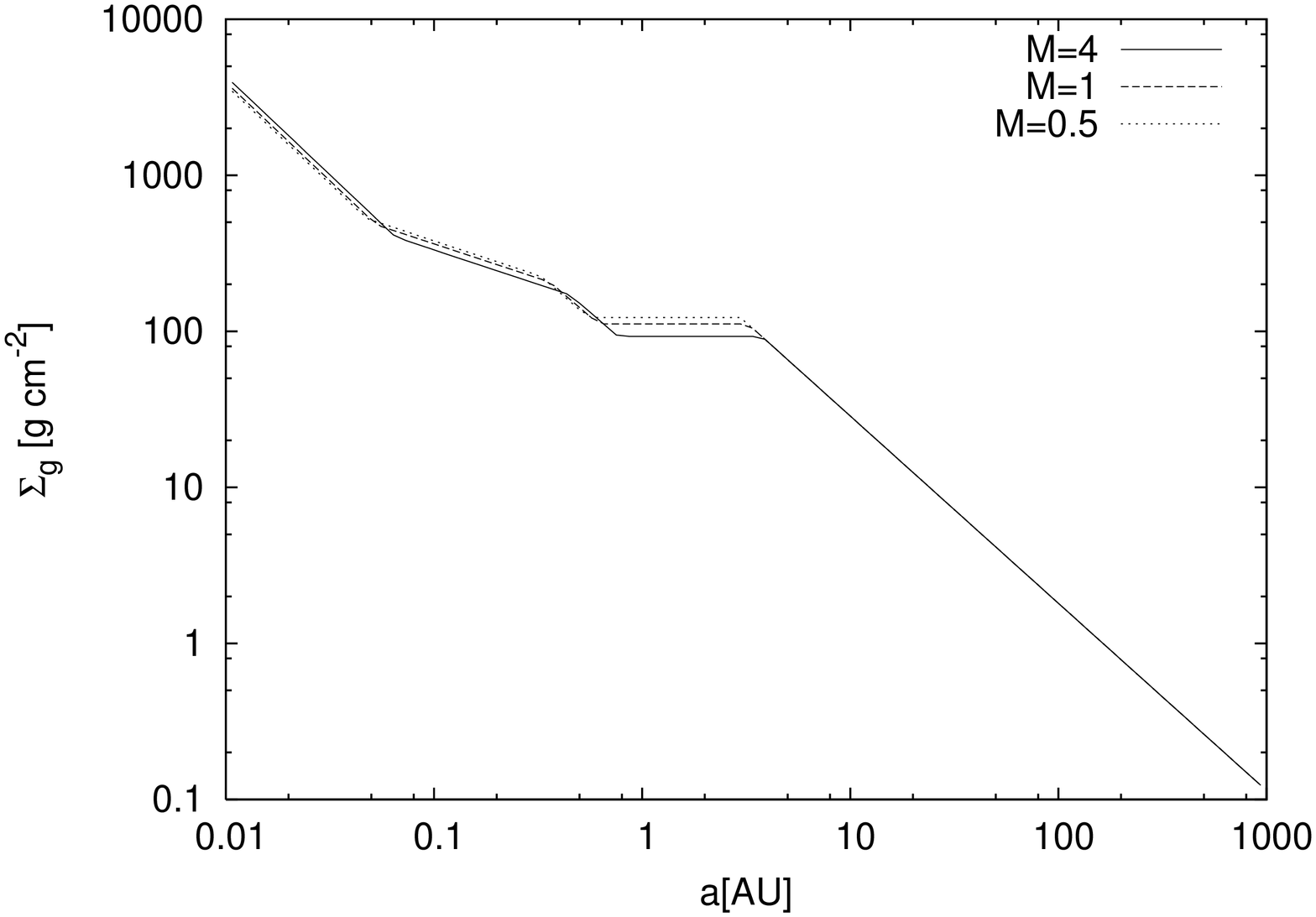}\includegraphics[angle=-90,width=8cm]{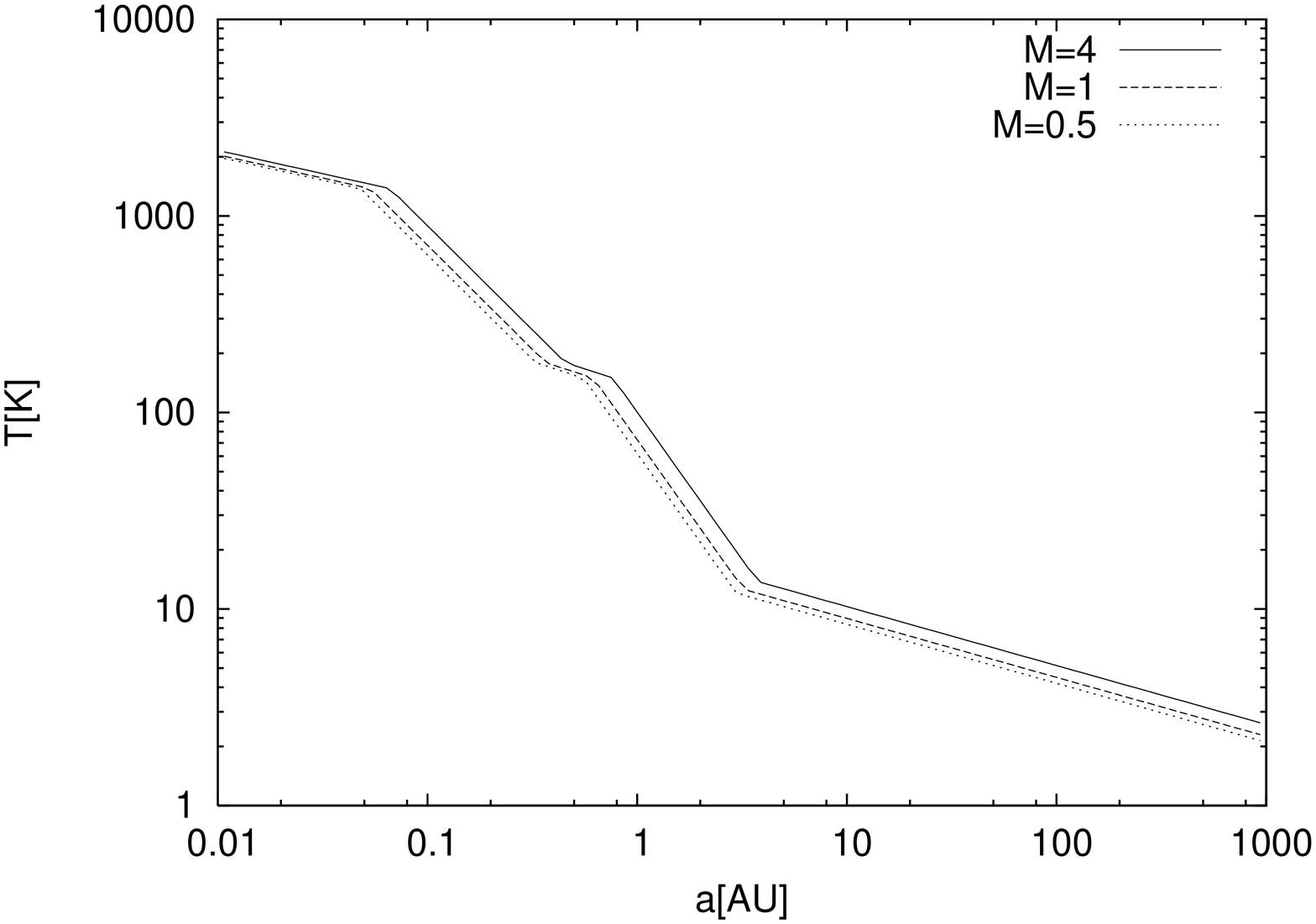}
\caption{The initial surface density (left panel) and temperature
  (right panel) as a function of distance from the star in the
  protoplanetary disk with an initial mass $M_0=0.1\ M_\odot$ and
  outer radius $R_0=1000\ \mathrm{AU}$. Different curves
  correspond to different values of the mass of the central star in solar
  masses as described by the labels.}
\label{f:gas}
\end{figure*}

To illustrate how the mass of the central star  influences the
formation of the planetesimal swarm, we follow the evolution of a
protoplanetary disk with an initial mass $M_0=0.1\ M_\odot$ and
initial outer radius $R_0=1000\ \mathrm{AU}$ for three values of
$M_\star$ (0.5,1 and 4 $M_\odot$). Figure \ref{f:gas} shows the
initial distributions of $\Sigma_\mathrm{g}$ and $T$ in the
midplane of the disk. In all three cases the gas is distributed
very similarly, with $\Sigma_\mathrm{g}$ dropping monotonically
from $\sim 4000\ \mathrm{g\ cm}^{-2}$ at 0.01 AU from the star to
$0.1\ \mathrm{g\ cm}^{-2}$ at the outer edge of the disk. The
changes of slopes in the distribution of $\Sigma_\mathrm{g}$
correspond to transitions between different powerlaws describing
the opacity of the disk matter in different temperature ranges
\citep{ruden91}. The distribution of temperature is qualitatively
very similar to the distribution of $\Sigma_\mathrm{g}$. In the
disk around a $1\ M_\odot$ star $T$ drops from $2500\ \mathrm{K}$
at $0.01\ \mathrm{AU}$ to a few Kelvins at the outer edge of the
disk. The evaporation temperature assumed for the dust grains in
our models ($1350\ \mathrm{K}$) is reached at $0.05 \
\mathrm{AU}$. Note that at a given radius $T$ increases as the
mass of the central star is increased. It is the result of the
increasing vertical component stellar gravity, due to which the
scale height of the disk is reduced.

\begin{figure}
\resizebox{\hsize}{!}{\includegraphics[angle=-90]{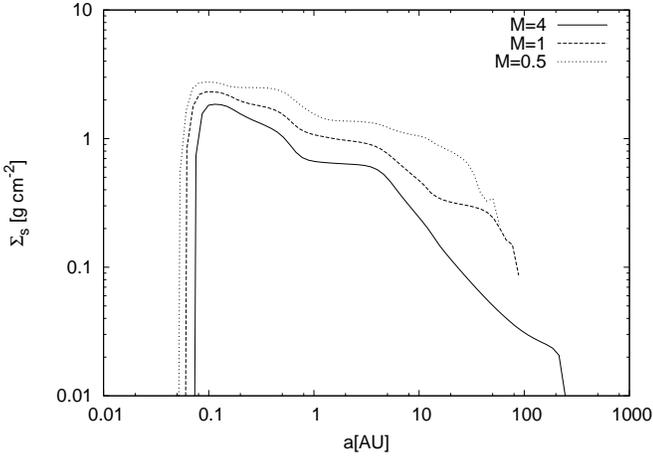}}
\caption{The surface density of solids as a function distance from the
  star in the protoplanetary disk with the same parameters as on
  Fig. \ref{f:gas} after $10^6 \mathrm{yr}$ from the beginning of its
  evolution. Different curves correspond to different
  values of the mass of the central star in solar masses as described by
  the labels.}
\label{f:sigend}
\end{figure}

Initially, the dust is well mixed with the gas, with the ratio
$\Sigma_\mathrm{s}/\Sigma_\mathrm{g}=6\times 10^{-3}$ being
constant everywhere in the disk. As the disk evolves, surface
density and temperature of gas slowly decrease due to accretion
and viscous spreading; however, the dust component evolves in 
quite a different way. The grains grow in size due to mutual
collisions and gain inward radial velocities due to the gas drag.
If they cross the evaporation radius, they sublimate and are
accreted onto the star on the viscous timescale as a vapour.
However, if their  growth time is shorter than the timescale of
inward migration, they manage to reach sizes of a few km {\it
before} reaching the evaporation radius. Their radial motions are
then stopped and the planetesimal swarm attains its final form.
Figure \ref{f:sigend} shows the distribution of planetesimals in our
model after $10^6$ yr from the beginning of its evolution. In all
three cases, the outer radius of the planetesimal swarm is much
smaller than the initial outer radius of the disk. The difference
is larger for models with smaller $M_\star$. This is because small
solid bodies evolving in disks around less massive stars gain
higher inward velocities and tend to migrate to smaller radius
before reaching km-sizes. The maximum velocity of the inward drift
can be estimated as
\begin{equation}
V_\mathrm{s}^\mathrm{max} \sim  \frac{C_\mathrm{s}^2}{a \Omega_\mathrm{K}}
\end{equation}
\citep{weiden77,kac2}. A change $M_\star$ influences it in two
opposite manners. On the one hand, $V_\mathrm{s}^\mathrm{max}$
increases with the mass of the star due to the increase in the
disk temperature . On the other hand, it decreases because of the
increase in the Keplerian velocity. Of these two competing factors,
the second one dominates, and as a result the
$V_\mathrm{s}^\mathrm{max}$ is a decreasing function of
$M_\star$.

Due to the inward migration of solids and their
confinement to much smaller radii, the final surface density of
planetesimals is increased locally within a factor of a few in
comparison with the initial value of $\Sigma_\mathrm{s}$. As this
effect is larger in disks around less massive stars, their final
planetesimal swarms tend to be more favourable to the formation
of giant planets.

\subsection{Minimum surface density}
To quantify the influence of $M_\star$ on the effectiveness
of giant planet formation from a planetesimal swarm, we introduce
the concept of the minimum surface density
$\Sigma_\mathrm{s,min}$. We define it as the minimum value of the
initial surface density of planetesimals $\Sigma_\mathrm{s,init}$
needed to form a Jupiter-mass (1M$_\mathrm{J}$) planet in less than the lifetime of
the protoplanetary disk $\tau_\mathrm{f}$. For $\tau_\mathrm{f}$
we adopt a value of $3\times 10^6\ \mathrm{yr}$.

\begin{figure}
\resizebox{\hsize}{!}{\includegraphics[angle=-90]{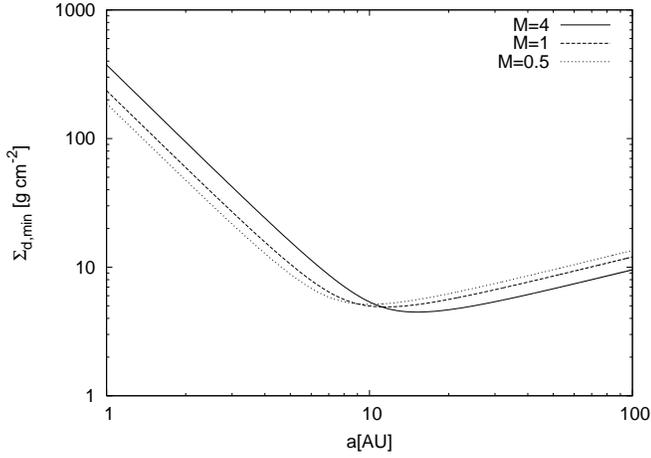}}
\caption{The minimum surface density of a planetesimal swarm
  needed to from a 1$M_\mathrm{J}$ planet in less than $3\times10^6$
  yr as a  function of distance from the central star. Different
  curves are obtained for different  masses of the central
  star, as labeled in  units of solar masses in the upper right corner.}
\label{f:sigmin}
\end{figure}

\begin{figure}
\resizebox{\hsize}{!}{\includegraphics[angle=-90]{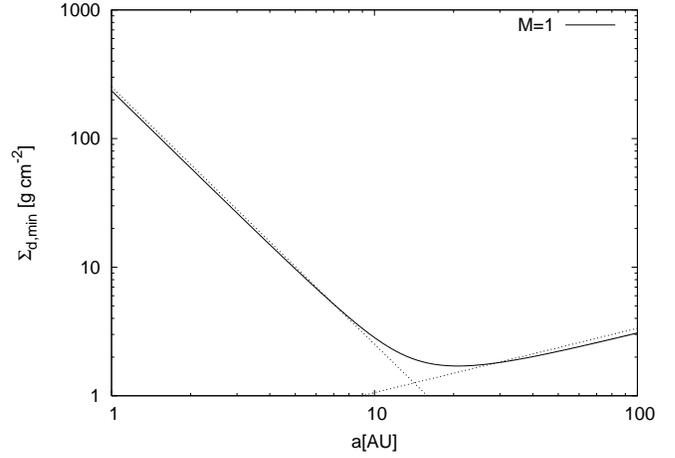}}
\caption{The minimum surface density of the planetesimal swarm
  needed to form a 1$M_\mathrm{J}$ planet in less than $3\times10^6$
  yr as a  function of distance from the central star, calculated
  with a constant value of $C_\mathrm{cap}=5$. The solid line represents the
  results of exact values obtained  numerically, while dotted
  lines represent analytical approximations given by
  Eqs. (\ref{eq:smin_low}) and (\ref{eq:smin_high}). The mass of the central
  star is $1\ M_\odot$.}
\label{f:sigmin_const}
\end{figure}

First, by solving the set of equations (\ref{eq:mc}),
(\ref{eq:sigmad}), and (\ref{eq:menv}) with different values of
$\Sigma_\mathrm{s,init}$, we determine $\Sigma_\mathrm{s,min}$ as
a function of distance from the star. The results are shown in Fig.
\ref{f:sigmin} for the same values of $M_\star$ as before. Close
to the star ($a<10\mathrm{AU}$), $\Sigma_\mathrm{s,init}$ is a
decreasing function $a$.  In this regime, the accreting
protoplanetary core rapidly accumulates all planetesimals in its
feeding zone and reaches the isolation mass $M_\mathrm{iso}$.
Afterwards, the accretion of planetesimals is negligible and the
planet grows mainly due to the accretion of gas. Upon integrating
Eq. (\ref{eq:menv}), we obtain the minimum isolation mass
needed to form a 1$M_\mathrm{J}$ planet within $\tau_\mathrm{f}$:
\begin{equation}
M_\mathrm{iso}\ge[M_\mathrm{J}^{1-c}+(c-1) \tilde{A}
\tau_\mathrm{f}]^{1/(1-c)},
\end{equation}
 where we assumed that the time needed for the core to reach
 $M_\mathrm{iso}$ is much shorter than $\tau_\mathrm{f}$. On the other hand,
 from the definition of $M_\mathrm{iso}$, we see that
\begin{equation}
M_\mathrm{iso}=\frac{(2 \pi)^{3/2}}{\sqrt{h}} a^3 \left(\frac{\Delta
  a}{R_\mathrm{H}}\right)^{3/2} M_\star^{-1/2}
  \Sigma_\mathrm{s,init}^{3/2}.
\end{equation}
Combining these two equations and neglecting the factor
$M_\mathrm{J}^{1-b}$, we obtain an analytical formula for
$\Sigma_\mathrm{s,min}$
\begin{equation}
\Sigma_\mathrm{s,min}=\frac{h^{1/3}}{2\pi} a^{-2}
\left(\frac{\Delta a}{R_\mathrm{H}}\right)^{-1} M_\star^{1/3}
[(c-1) \tilde{A}  \tau_\mathrm{f}]^{2/3(1-c)}, \label{eq:smin_low}
\end{equation}
  (\ref{eq:smin_low})
  which for $a~<~10$ AU agrees very well with the exact values of
$\Sigma_\mathrm{s,min}$ on the descending branch of the curve
$\Sigma_\mathrm{s,min}(a)$ (see Fig. \ref{f:sigmin_const}). Note
that in this regime the quantity $\Sigma_\mathrm{s,min}$ is an
increasing function of the stellar mass. This behaviour is due to
the fact that, while we increase $M_\star$, the Hill sphere shrinks,
and as a result the isolation mass decreases (provided, of course,
that $\Sigma_\mathrm{s,init}$ stays constant).

For sufficiently large radii $(a>10\ \mathrm{AU})$,
$\Sigma_\mathrm{s,min}(a)$ changes its slope and becomes an
increasing function of $a$. In this regime, the time scale of
accretion of the planetesimals onto the core is larger than the
lifetime of the disk and becomes the main factor to determine
$\Sigma_\mathrm{s,min}$. Consequently, the surface density of
planetesimals in the feeding zone never drops much below its
initial value. To describe the formation of a planet analytically
under these conditions, we divide the whole process into two
phases. During the first phase the planet exclusively due to
the accretion of planetesimals grows . We assume that the surface density
of planetesimal swarm $\Sigma_\mathrm{s}$ does not change in time,
because the core only accumulates  a negligible fraction of solids
present in the feeding zone. From Eq. (\ref{eq:mc}) we obtain
\begin{equation}
\frac{\mathrm{d} M_\mathrm{p}}{\mathrm{d} t} = A_\mathrm{c}
M_\mathrm{p}^{2/3}, \label{eq:md}
\end{equation}
where
\begin{equation}
A_\mathrm{c}=\left(\frac{3 \pi^2}{4}\right)^{1/3} \sqrt{G} C_1 C_\mathrm{cap} \frac{1}{(h
  \rho_\mathrm{c})^{1/3}} \frac{M_\star^{1/6}}{\sqrt{a}} \Sigma_\mathrm{s,init}
\label{eq:A}
\end{equation}
is a constant factor (we neglect the changes of $C_\mathrm{cap}$
in time, and $\rho_\mathrm{c}$ in the above expression  denotes
the density of the core). During the second phase the planet grows
exclusively due to the accretion of gas at a rate described by Eq.
(\ref{eq:menv}). This phase begins when the planet reaches the
mass $M_\mathrm{p,int}$ for which the accretion rate of gas is
equal to the accretion rate of  planetesimals. From
Eqs.(\ref{eq:menv}) and (\ref{eq:md}), we get
\begin{equation}
M_\mathrm{p,int}=\left(\frac{A_\mathrm{c}}{\tilde{A}}\right)^{3/(3c-2)}
\label{eq:mint}
\end{equation}
Integrating Eqs. (\ref{eq:md}) and (\ref{eq:menv}) over time yields
the lengths of both phases $\Delta t_1$ and $\Delta t_2$:
\begin{equation}
\Delta t_1 =\frac{3 M_\mathrm{p,int}^{1/3}}{A_\mathrm{c}},
\label{eq:t1}
\end{equation}
and
\begin{equation}
\Delta t_2 =\frac{ M_\mathrm{p,int}^{1-b}}{(b-1)\tilde{A}}
\label{eq:t2}
\end{equation}
Combining Eqs. (\ref{eq:A}), (\ref{eq:mint}), (\ref{eq:t1}), and
(\ref{eq:t2}) with the condition that $\Delta t_1 + \Delta t_2 \leq
\tau_\mathrm{f}$, we get the equation for $\Sigma_\mathrm{s,min}$ in that regime:
\begin{equation}
\Sigma_\mathrm{s,min}=\left(\frac{4}{3 \pi^2}\right)^{1/3}
         \frac{(h \rho_\mathrm{c}])^{1/3}}{\sqrt{G} C_1 C_\mathrm{cap}}
         \tilde{A}^{\frac{1}{3(1-c)}}
         \left[\frac{c-1}{3c-2} \tau_\mathrm{f} \right]^{\frac{3c-2}{3(1-c)}}
         \frac{\sqrt{a}}{M_\star^{1/6}}
\label{eq:smin_high}
\end{equation}
 With $C_\mathrm{cap}=1.2$, the above formula satisfactorily
reproduces the rising branch of curve
$\Sigma_\mathrm{s,min}(a)$ (see Fig. \ref{f:sigmin_const}). Note
that the ``best-fit'' value of $C_\mathrm{cap}$ is consistent with
the maximum and minimum values of this parameter used in our
numerical calculations ($1$ and $5$, respectively).
%We also tested that our formula agrees
%very well with values from the numerical calculations of
%$\Sigma_\mathrm{s,min}$, in which constant factor $C_\mathrm{cap}$
%was used
At large distances from the star, $\Sigma_\mathrm{s,min}$ is a
decreasing function of $M_\star$, because increasing the mass of
the star causes the accretion rate of planetesimals to
decrease (see Eq. \ref{eq:md}).

 As we see, the minimum surface density of the planetesimal
swarm required for the formation of a gas giant planet in a time
shorter than the lifetime of protoplanetary disk is a complicated
function of $M_\star$. In a given planetesimal swarm such a planet
forms more easily around less massive star if its orbital radius
is smaller than $\sim$ 10 AU, and around a more massive star if
its orbital radius is larger than $\sim$ 10 AU.
%At small distances from the
%star it is  from the same around stars with different masses is
%more (less) difficult around more (less) massive star. For larger
%distances the relation is inverted.

\subsection{Grid of models}
\label{s:grid} 
\begin{figure}
\resizebox{\hsize}{!}{\includegraphics{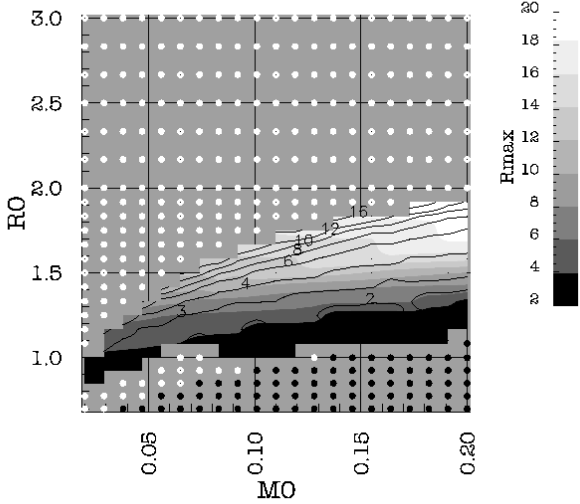}} \caption{
The plane of initial parameters of protoplanetary disk models
[$M_0$, $\log R_0$]. Minimum and maximum distance from a $1
M_\odot$ star at which the formation of a 1$M_\mathrm{J}$ planet
is possible within $3\times10^6$ yr is indicated by contours and
a grey scale, respectively. White circles indicate disk models in
which the surface density of the planetesimal swarm is everywhere
lower than the critical value for planet formation. Black circles
indicate disks in which all solids are accreted onto the star.}
\label{f:1.0}
\end{figure}

\begin{figure}
\resizebox{\hsize}{!}{\includegraphics{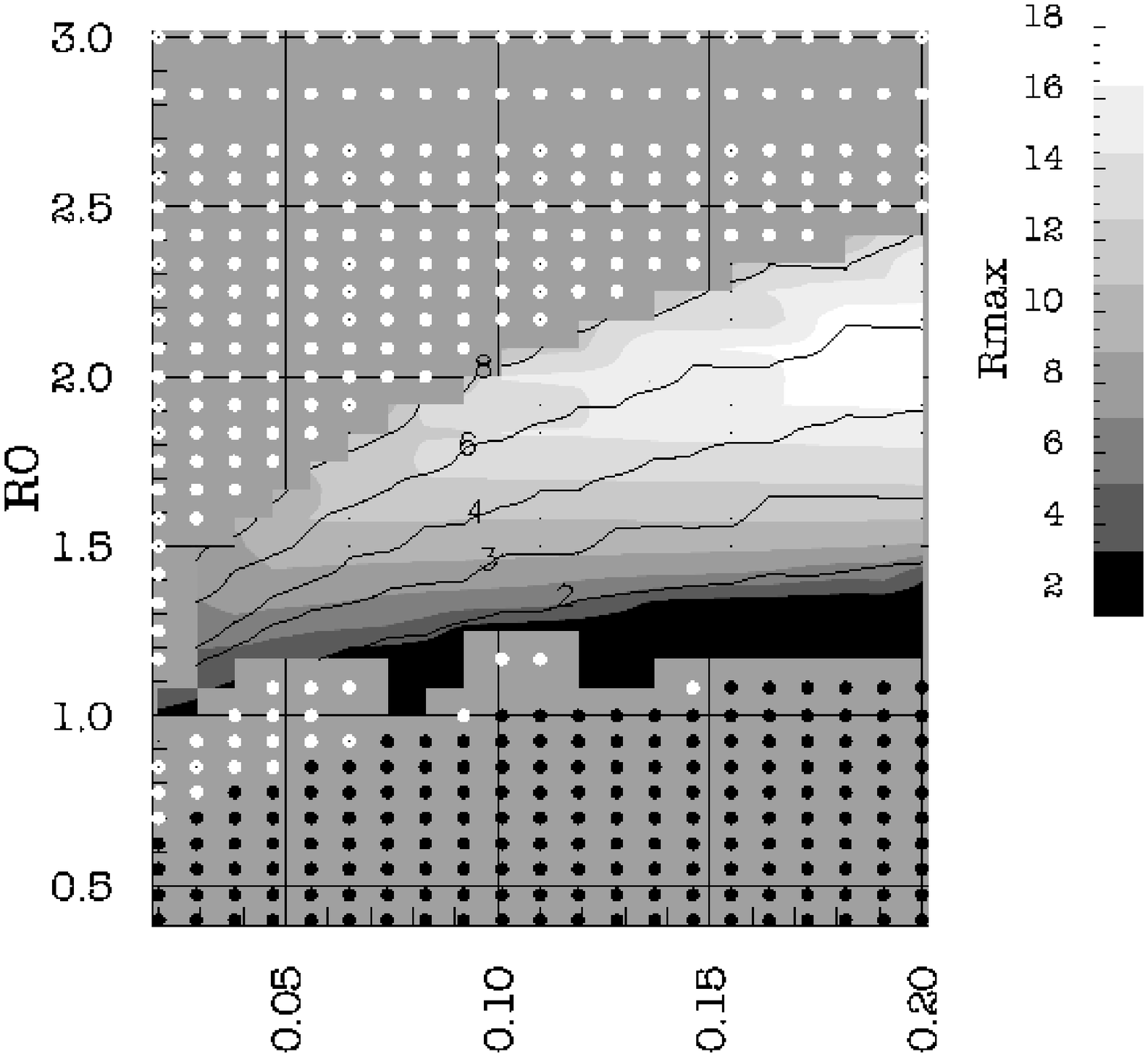}}
\caption{Same as Fig. \ref{f:1.0} but for the central star with a mass
  of $0.5 M_\odot$.}
\label{f:0.5}
\end{figure}

\begin{figure}
\resizebox{\hsize}{!}{\includegraphics{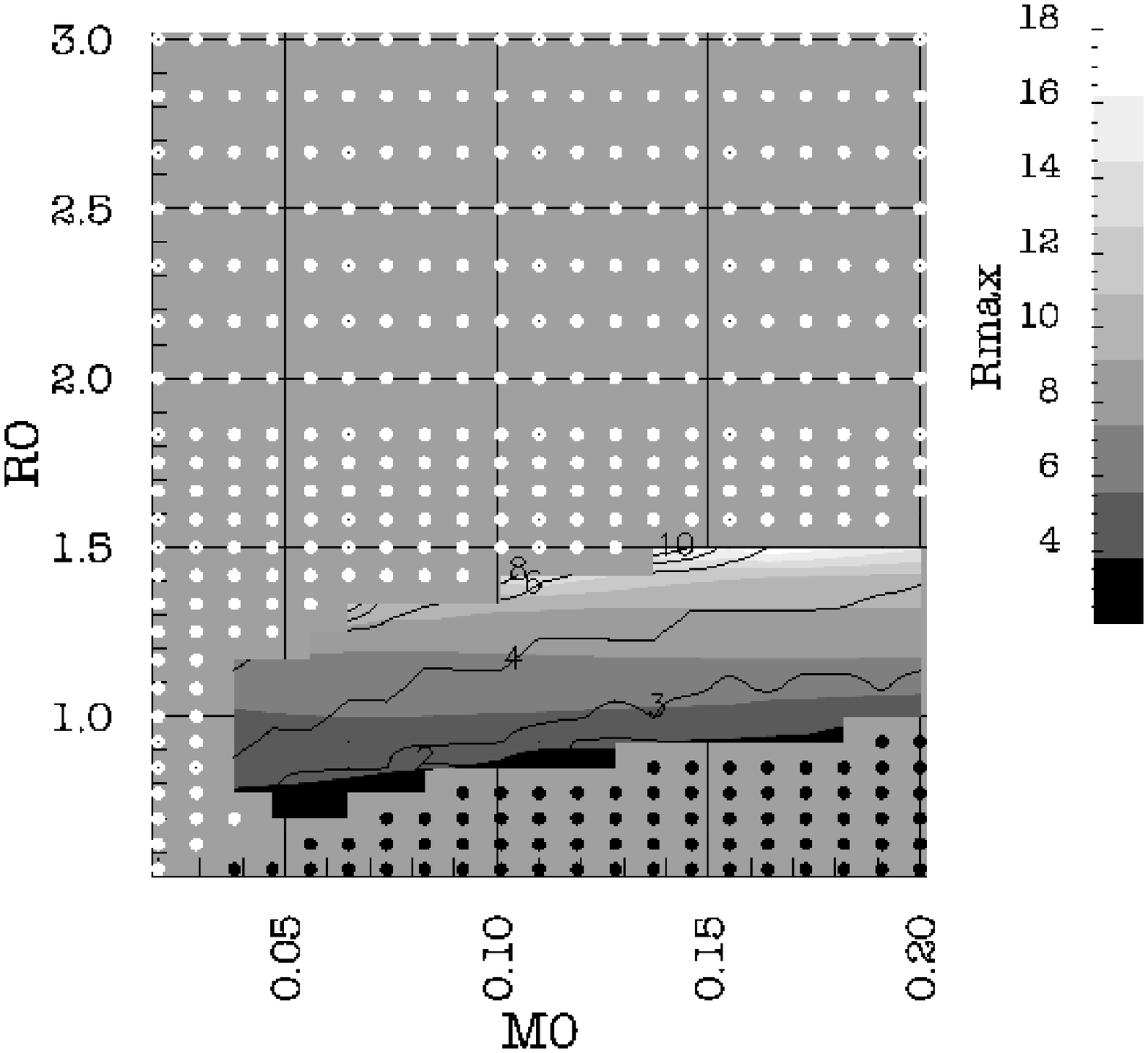}}
\caption{Same as Fig. \ref{f:1.0} but for the central star with a mass
  of $4 M_\odot$.}
\label{f:4.0}
\end{figure}

The results of the last two subsections allow us to
investigate the influence of $M_\star$ on the whole process of
giant planet formation for a broad set of models of protoplanetary
disks. We calculate the grid of models similar to the one
described in Sect. \ref{s:exam}, but with different values of the
initial disk mass $M_0$ and outer radius $R_0$. To cover the range
of masses and sizes of disks observed in nature, we choose $M_0$
in the range of 0.02 to 0.2 $M_\odot$. The range of $R_0$ is
adjusted for every metallicity so that all models in which
formation of giant planets is possible could be accounted for. For
the viscosity coefficient $\alpha$ we chose a value of 0.001
\citep{papaloizou03}. We follow each model until all solids are in
the form of planetesimals or are accreted onto the star. Then, we
evaluate every model with planetesimals to determine whether the
surface density of planetesimals exceeds the minimum surface
density $\Sigma_\mathrm{s,min}$ anywhere in the disk. Models with
this property are labeled as \textit{planet bearing}. For each
such model we determine the minimum and maximum distances from the
star at which the surface density of the planetesimal swarm is
larger than $\Sigma_\mathrm{s,min}$. The results obtained for
different values of the stellar mass are shown in Fig.
\ref{f:1.0}-\ref{f:4.0}.

The area of the region occupied on the $[M_0,R_0]$ plane by the
planet bearing disks is clearly anticorrelated with the mass of the
central star. As we have shown in Sect. \ref{s:exam}, solid grains
gain higher inward velocities in disks around less massive stars,
and the resulting planetesimal swarms have higher surface
densities. We see that reducing the mass of the central star
increases the maximum $R_0$ for which planet formation is possible
in disks with the same initial mass $M_0$.

 We can also see that in a given disk the inner edge of the
planet-bearing region moves inward as we decrease the mass of the star
(for $M_\star=0.5 M_\odot$ its radius is on the average equal to
$\sim0.55$ that for $M_\star=1 M_\odot$).  This effect is mainly
caused by the larger surface density of planetesimal swarms produced
by disks around less massive stars. Differences in
$\Sigma_\mathrm{s,min}$, while appreciable, are much less important
(when $M_\star$ is reduced by a factor of 2, $\Sigma_\mathrm{s,min}$
drops by 20\% only; see Eq.  \ref{eq:smin_low}). The outer edge of the
planet-bearing region also moves closer to the star (for $M_\star=0.5
M_\odot$ its radius is on the average equal to $\sim0.75$ that for
$M_\star=1 M_\odot$).  This is because in most cases it coincides with
the outer edge of the planetesimal swarm, which is more compact around
less massive stars. The difference in the minimum surface density
$\Sigma_\mathrm{s,min}$ also tends to decrease the outer radius of the
planet-bearing region, but it is again a second-order factor.

Generally, our model predicts that giant planets tend to form
at tighter orbits around less massive stars, and wider orbits
around more massive stars. However, at least in some cases their
locations may be influenced by the effects of migration. We return
to this point in Sect. \ref{s:conc}.

%Moreover, provided that the initial population of the $[M_0,R_0]$
%plane does not depend too strongly on $M_\star$, the rate of
%planet occurrence should be anticorrelated with the mass of the
%central star.

\subsection{Metallicity relation}
One of the main results of extrasolar planet searches is the
discovery that planet-bearing stars tend to have higher
metallicities than field stars \citep{santos00,debra03}. That
correlation can be easily explained within CAGCM. In this scenario,
the formation time of giant planets decreases with increasing
surface density of the planetesimal swarm (see Eqs. (\ref{eq:smin_low})
and (\ref{eq:smin_high})), which in turn is an increasing function
of the original metal content of the protoplanetary disk.
Consequently, the giant planets form more easily in disks with
higher metallicities. \citet{kac05} calculated the rates of giant-planet
occurrence in disks with different metallicities around
stars with mass $M_\star=1 M_\odot$. Their approach to the
evolution of solids and to the formation of giant planets was the same as
the one used in this paper.  They were able to reproduce the
observational correlation for disk models with viscosity parameter
$\alpha = 10^{-3} - 10^{-2}$. Herein we extend their calculations
onto disks around stars with various masses.

The change in the disk metallicity influences the processes
leading to the formation of planets in two ways. First, it
changes the structure of the gaseous disk by changing the opacity
in the disk. In our models we scaled the opacity by a constant
factor $Z$ equal to the metallicity of the disk expressed in solar
units. This approach is justified by the fact that the opacity in
protoplanetary disks is mainly due to dust grains and molecules.
Second, the primordial metallicity of the disk determines the
initial ratio of dust-to-gas surface densities. In our models this
ratio is initially independent of the distance from the star and
is equal to
\begin{equation}
\Sigma_\mathrm{s}/\Sigma_\mathrm{g}=6\times 10^{-3} Z \mathrm{.}
\end{equation}

We compute grids of models similar to those described in Sect.  \ref{s:grid}
for eight different values of $Z$ distributed between 0.2 and 3. Additionally,
we check the gravitational stability of the corresponding gaseous disk for every
value of $[M_0, R_0]$ . In some cases the value of the Toomre parameter
\begin{equation}
  Q=\frac{C_\mathrm{S} \Omega_\mathrm{K}}{\pi G \Sigma},
\end{equation}
where $\Sigma=\Sigma_\mathrm{g}+\Sigma_\mathrm{s}$, drops below 1
in the outer region of the disk, which means that they are unstable
with respect to axisymmetric modes. We assume that such region
fragments  and giant planets are formed there on a very short
time scale, consuming and/or dispersing the unstable part of the
disk. In such cases we use modified initial values of $M_0$ and
$R_0$, which correspond to the mass and outer radius of the stable
part of the original disk. In principle, some ''special treatment"
should also be applied to disks with $1\le Q \lesssim 1.3$, which
develop spiral arms and for a while evolve due to gravitational
rather than viscous angular momentum transfer \citep[see e.g.][]{laughlin}.
However, since the efficiency of gravitational transport quickly
decreases with the ratio $M_0/M_\star$ and it might
become significant in our study only for the most massive disks around 0.5
$M_\odot$ stars, we decided to neglect this effect altogether.

Following the procedure described by \citet{kac05}, for every $Z$
we calculate the area $A_\mathrm{p,5}$ of the region occupied on
the $[M_0, \log R_0]$ plane by disks that form planets at
distances smaller than 5 AU from the central star. The last
restriction comes from the fact that currently we know only one
extrasolar planet  on a larger orbit\footnote{see The Extrasolar
Planets Encyclopedia at\\ http://www.obspm.fr/encycl/encycl.html}
-- 55~Cnc~d \citep{marcy02}. A measure of the rate of planet
occurrence can be defined as
\begin{equation}
P_\mathrm{p}=\frac{A_\mathrm{p,5}}{C},
\end{equation}
where the normalization factor $C$ is chosen in such a way as to
reproduce the observed value of $P_\mathrm{p}$ for $Z=3$ and
$M_\star=1\ M_\odot$.

\begin{figure}
\resizebox{\hsize}{!}{\includegraphics[angle=-90]{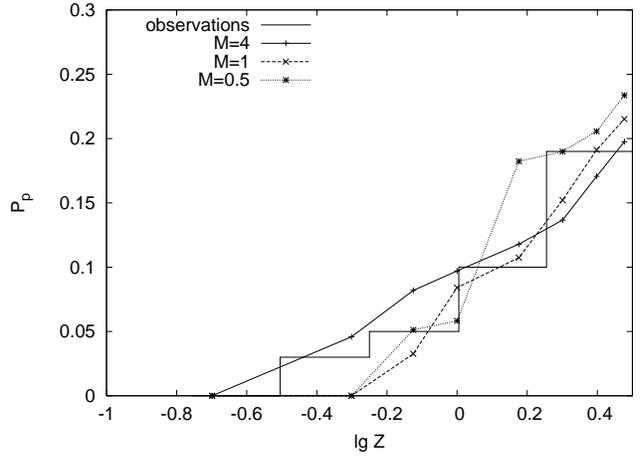}}
\caption{The rate of planet occurrence as a function of the
primordial metallicity of protoplanetary disks. Different lines are
obtained for models with different masses of the central star, as
labeled in solar units in the upper left corner. The histogram
shows the observational data compiled by \cite{debra03}.}
\label{f:P_p}
\end{figure}

The results are presented in Fig. \ref{f:P_p}. It shows the rate
of planet occurrence as a function of disk metallicity for three
values of $M_\star$. As expected, $P_\mathrm{p}$ is an increasing
function of $Z$. The minimum value of $Z$ below which no giant
planets are formed at orbits smaller than 5 AU decreases with the
mass of the central from $Z_\mathrm{min}\approx0.6$ for
$M_\star=0.5 M_\odot$, to $Z_\mathrm{min}\approx0.2$ for
$M_\star=4 M_\odot$.

 We see that for $Z$ smaller than $\sim$0.2, $P_\mathrm{p}$ is
an increasing function of $M_\star$. This is because most disks in
which planets formation would be possible around less massive
stars have outer parts that are gravitationally unstable, and the amount of
solids present in their stable parts is too low to enable
subsequent formation of giant planets according to CAGCM.
%For metal-rich stars it is not longer valid.
However, as $Z$ increases, smaller and smaller disks become
planet-bearing for every $M_0$, and the percentage of
gravitationally unstable disks in which formation of giant planets is
not possible  decreases.
%value of the initial mass of the diskthe planet bearing region on $[M_0, \lg R_0]$
%diagram contains. As a result the larger number entirely stable
%and favourable for the formation of giant planets disks appear.
Consequently, the factors promoting the formation of a giant planet
around less massive stars as described in previous sections become
important, and $P_\mathrm{p}$ changes into a decreasing function
of $M_\star$.

\section{Conclusions}
\label{s:conc}
Based on a simple approach to the evolution of solids in
protoplanetary disks, we investigated the influence of the mass of the
central star on the formation of giant planets.
We showed that due to the more efficient redistribution of
solids the planetesimal swarms around less massive stars tend to
have higher surface densities. Next, we derived the minimum
surface density of the planetesimal swarm needed to enable
formation of a giant planet within the lifetime of the
protoplanetary disk, and we found that at distances from the star
smaller than $\sim\!10$~AU it increases with the stellar mass.
%These two factors
%make the formation of giant planets around less massive stars
%easier, but the first one is much more important than the second.
Farther away from the star the minimum density becomes
anticorrelated with the mass of the star. However this effect is
offset by the anticorrelation mentioned already between the mass
of the star and the surface density of the planetesimals.

These two effects determine the set of initial parameters
characterising protoplanetary disks that are capable of giant
planet formation within the core accretion - gas capture scenario.
We showed that this set is larger for less massive stars.
This means that the percentage of stars with massive planets
should increase with decreasing stellar mass (at least in the
range 0.5 $M_\odot$ -- 4 $M_\odot$). However, as discussed below,
in the currently accessible range of orbital radii ($<$ 5 AU), the
situation is not all that clear.

Based on the sets obtained for different metallicities, we
determined the occurrence rate of planets with orbits smaller than
$5$~AU as a function of the mass and metallicity of the star. We took
into account the fact that  the outer region of the
disk is gravitationally unstable in some models. Such regions are located farther
than $5$~AU from the central star, and planets formed there by
disk fragmentation are not included in our occurrence rate.
However, their presence reduces the amount of solid material
available for the formation of planetesimals. For less massive
stars this effect is so strong that it overcomes factors promoting
planet formation, so that for metal-poor disks the rate of planet
occurrence decreases with the mass of the central star. As a
result, the minimum metallicity at which giant planets can form at
orbits smaller than $5\ \mathrm{AU}$ decreases from $\sim 0.6$ for
stars with masses of $0.5 M_\odot$ to $\sim 0.2$ for $4 M_\odot$.

In the metal-rich regime the percentage of entirely stable disks
in which formation of giant planets is possible is larger, and
stable regions of partly unstable disks contain enough solids to
produce planetesimal swarms capable of giant planet formation.
Consequently, both factors promoting planet formation around less
massive stars are in play, and a clear anticorrelation between
the stellar mass and planet occurrence rate is observed. At
the same time, our model does not account for the presence of giant
planets around metal-poor stars. This may be due to the fact that
we do not include planets that have formed beyond 5 AU and later
migrated inward. Such an assumption is valid as long as the number
of these planets is small compared to the number of planets that
have formed within 5 AU. However, as we move to lower
metallicities, the percentage of giant planets with silicate cores
decreases (the silicates simply become too scarce), while the
percentage of planets forming from ice grains increases. Thus, in
metal-poor systems the number of planets with ice cores that
migrated from large orbits can become a large fraction of planets
at orbits smaller than 5 AU.

\begin{figure}
\resizebox{\hsize}{!}{\includegraphics[angle=-90]{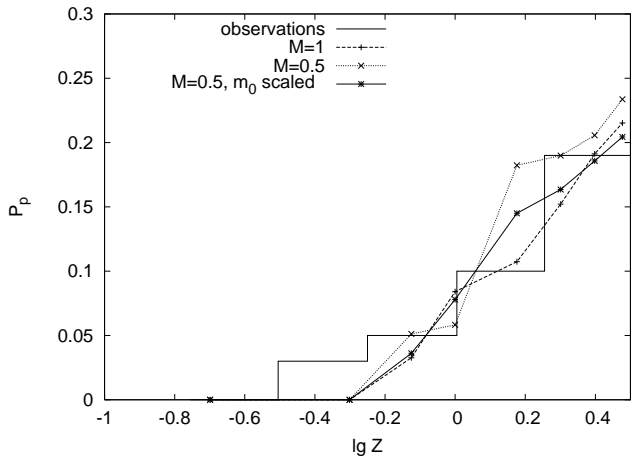}}
\caption{The rate of planet occurrence as a function of the
primordial metallicity of protoplanetary disks. Disks around  stars with masses
of $1 M_\odot$ are represented by a solid line. Disks around stars with
masses of $0.5 M_\odot$  are represented by dotted and dashed lines. In
the first case the range of initial masses of the disks is the same as
in the case of solar type stars, while in the second it was scaled according
to the mass of the central star.}
\label{f:rout_sc_t}
\end{figure}

Obviously, our description of the evolution of solids is very
simplified. The basic underlying assumptions like the single-size
distribution of solid grains or the neglect of planet migration
already have been discussed by \citet{kac4} and \citet{kac05}. The main
additional assumption introduced in the present paper is the
independence of the initial parameters of protoplanetary disks on
the mass of the central star. While admittedly {\it ad hoc}, it
seems to be better than the one adopted by \citet{laughlin04}, who
scaled their initial surface density of planetesimals linearly
with the mass of the star. They did not take into account the
antecedent evolution of solids leading to the formation of
planetesimal swarms, and concluded that the formation of giant
planets around low-mass stars is difficult. Recent
observations suggest that masses of protoplanetary disks do not
strongly depend on masses of the central stars \citep{guilloteau}.
Nevertheless, to investigate the influence of our assumption, we
performed additional set of calculations with a mass of the central
star of $0.5 M_\odot$ and with the initial masses of disks scaled by
factor of $0.5$. The results are shown in Fig. \ref{f:rout_sc_t}.
In this case the probability of finding a planet does not seem to
depend strongly on the mass of the central star,  which is true
for the whole range of metallicities we have considered. Still,
our models show that the evolution of solids leading to the
formation of planetesimal swarms is a vital factor facilitating
the formation of giant planets, whose role should be particularly
clear for low-mass stars.

Our models of gaseous disks do not reproduce recent
observations by \citep{muzerolle}, which show that the accretion
rate in protoplanetary disks increases with the mass of the
central star. However, in the mass range considered here this
dependence is very weak, and for a given value of stellar mass the
spread in accretion rates reaches two orders of magnitude. In our
opinion these data do not invalidate our basic assumption that
initial disk parameters do not depend on the mass of the star. We
also assumed that heating by stellar radiation is negligible,
whereas at least in some cases it can be a dominant source of
energy in the outer regions of the disk (more efficient than the
turbulent dissipation). As such, it may substantially change the
structure of the disk and the radial velocities of solids.
Currently we are working on models that will take these effects
into account.

% From the observational point of view we lack any
%evidences for way in which initial masses and sizes of protoplanetary
%disks scale with the mass of the central star.
\begin{acknowledgements}
This project was supported by the German Research Foundation (DFG)
through the Emmy Noether grant WO 857/2-1 and the European
Community's Human Potential Programme through the contract
HPRN-CT-2002-00308, PLANETS. KK and MR acknowledge  support
from the grant No. 1 P03D 026 26 from the Polish Ministry of
Science.
\end{acknowledgements}

\bibliography{metallicity}
\bibliographystyle{aa}

\end{document}